\documentclass[conference]{IEEEtran}
\usepackage{amsmath,amsfonts}
\usepackage{algorithmic}
\usepackage{algorithm}
\usepackage{array}
\usepackage[caption=false,font=footnotesize,labelfont=rm,textfont=rm]{subfig}
\usepackage{textcomp}
\usepackage{stfloats}
\usepackage{url}
\usepackage{verbatim}
\usepackage{graphicx}
\usepackage{cite}
\usepackage{bm}
\usepackage{color}
\begin{document}
\title{Graph Neural Network for Location- and Orientation-Assisted mmWave Beam Alignment}



\author{
    \IEEEauthorblockN{Yuzhu Lei, Qiqi Xiao, Yinghui He, Guanding Yu$^*$}
    \IEEEauthorblockA{ College of Information Science and Electronic Engineering, Zhejiang University, Hangzhou 310027, China}
    \IEEEauthorblockA{Email: leiyuzhu@zju.edu.cn, xiaoqiqi@zju.edu.cn, 2014hyh@zju.edu.cn, yuguanding@zju.edu.cn} 
}

\maketitle

\begin{abstract}
In massive multi-input multi-output (MIMO) systems, the main bottlenecks of location- and orientation-assisted beam alignment using deep neural networks (DNNs) are large training overhead and significant performance degradation. This paper proposes a graph neural network (GNN)-based beam selection approach that reduces the training overhead and improves the alignment accuracy, by capitalizing on the strong expressive ability and few trainable parameters of GNN. Due to the spatial correlation, the channels of beams with similar beam directions are generally correlated. Therefore, we establish a graph structure according to the angular correlation between beams and use GNN to capture the channel correlation between adjacent beams, which helps accelerate the learning process and enhance the beam alignment performance. Compared to existing DNN-based algorithms, the proposed method requires only 20\% of the dataset size to achieve equivalent accuracy and improves the Top-1 accuracy by 10\% when using the same dataset.

\end{abstract}


\section{Introduction}
Beamforming is an attractive solution for enabling millimeter wave (mmWave) communications that utilizes the high antenna gain to compensate for the excessive path loss by establishing a directional link between the transmitter (TX) and receiver (RX) \cite{ref12,ref17}. However, the beamforming process also introduces additional latency and overhead to mmWave communications. In \cite{ref1}, the authors propose a codebook-based beamforming approach to simplify the beam alignment process. Consequently, transceivers select the appropriate precoder and combiner from pre-defined codebooks to transmit and receive signals on the line-of-sight (LOS) path or strong non-line-of-sight (NLOS) paths \cite{ref13}.

Traditional exhaustive search method obtains high accuracy by scanning through all combinations of precoders and combiners, resulting in unacceptable latency and overhead. Hierarchical beam search (HBS) uses hierarchical codebooks to reduce search cost by gradually narrowing the scanning range according to the feedback from counterpart \cite{ref16}. However, the low antenna gain and insufficient spatial resolution of wide beams used in HBS not only limit the service range but also lead to incorrect decisions, resulting in a degradation in accuracy. 

Utilizing context information (CI), especially the location and orientation of the user equipment (UE), is currently the main trend to reduce the overhead and latency of beam alignment \cite{ref14,ref15,ref18}. 
Besides, through training historical data with deep learning (DL) methods, data features can be extracted to effectively mitigate the high complexity and latency problems in beam management \cite{ref11}.
In \cite{ref2,ref3}, deep neural networks (DNNs) are used for classification, which take the location and orientation of UE as input and output the optimal probability of each beam pair for beam selection. The DNN-based methods significantly outperform the HBS, lookup table method, K-nearest neighbor (KNN)-based method, and support vector machine (SVM)-based method \cite{ref2,ref3,ref4}. However, the performance of DNN-based methods drops dramatically as the size of antenna arrays increases \cite{ref4}. Besides, due to frequent environmental changes, the training process needs to be continuously executed. High training overhead leads to increased energy consumption and impedes the rapid deployment of DNN-based methods, which conflicts with the low energy requirements anticipated in next-generation networks \cite{ref5}. How to deal with limited training samples and maintain high accuracy remains a challenge.

In \cite{ref6}, the authors propose using transfer learning (TL) to improve the performance of DNN-based methods when only a small dataset is available, but its effectiveness is influenced by the similarity between datasets and the parameter freezing operations. 
The authors of \cite{ref7} propose a deep regularization method that exploits the temporal correlation features of data and achieves equivalent performance with fewer training samples, indicating the potential of exploiting data correlation in reducing training overhead and improving performance. In \cite{ref8}, the authors propose a DL-based low overhead beam selection method, which uses convolutional layers to utilize the spatial correlation of beam domain image. However, the spatial correlation between beams is not fully utilized.

In this paper, we propose to fully leverage the spatial correlation between beams to construct an efficient and accurate DL model for location- and orientation-assisted beam alignment. Compared with DNN, graph neural network (GNN) has the advantages of good expression ability and few trainable parameters. Therefore, we propose to establish a graph structure according to the angle correlation between beams and use GNN to capture the correlation between adjacent beams, in order to speed up the learning process and improve the beam alignment accuracy. To evaluate the proposed method, we consider the DNN-based beam alignment in \cite{ref4} and the DNN-based beam alignment with TL in \cite{ref6} as benchmarks. Simulation results indicate that the proposed method can significantly reduce training overhead and enhance beam alignment performance.

\section{System Model}
We consider a 3D mmWave communication system consisting of one TX and one RX equipped with uniform linear arrays (ULAs), while the GNN-based beam alignment can be straightforwardly adapted to the multi-user scenario. We assume the ULAs of TX and RX are placed in the x-y plane and have $N_t$ and $N_r$ ideal isotropic antenna elements separated by $\lambda/2$, respectively. To simulate the rotation of UE, we also consider the ULA of RX rotates around the z-axis, with the orientation provided by the rotation angle.

\subsection{Channel Model}
For simplicity, the TX is located at $\bm p_t=(0,0,0)$ with a fixed orientation $\alpha_t$, and the RX is placed at a random location $\bm p_r=(x_r,y_r,z_r)$ within the service area, with a random orientation $\alpha_r\in[0,2\pi)$. The channel matrix $\bm H$ between TX and RX with one LOS path and $L$ NLOS paths is modeled as
\begin{equation}
\label{deqn_ex1}
\bm H=\sum_{l=0}^{L}\sqrt{\rho_{l}}e^{j\vartheta_l}\bm a_r(\phi_{r,l},\theta_{r,l})\bm a_t^H(\phi_{t,l},\theta_{t,l}),
\end{equation}
where $\rho_l$ and $\vartheta_l$ are the path attenuation and phase of the $l$-th path, $\phi_{r,l}$ and $\theta_{r,l}$ are the azimuth and elevation of the angle of arrival (AOA), $\phi_{t,l}$ and $\theta_{t,l}$ are the azimuth and elevation of the angle of departure (AOD). The antenna array response of the TX can be described as
\begin{equation}
\label{deqn_ex2}
\begin{aligned}
\bm a_t(\phi_{t,l},\theta_{t,l})=\frac{1}{\sqrt{N_t}}[&1,e^{j\pi \sin(\theta_{t,l})\cos(\phi_{t,l})},...,\\
&e^{j\pi(N_t-1) \sin(\theta_{t,l})\cos(\phi_{t,l})}]^T.
\end{aligned}
\end{equation}
The antenna array response of the RX is analogously defined.

\subsection{Beam Codebook}
For the convenience of analysis, we employ the discrete Fourier transform (DFT) codebooks. The codebooks of the TX and RX are denoted as $\mathcal{U}=\{\bm u_1,...,\bm u_{N_t}\}$ and $\mathcal{V}=\{\bm v_1,...,\bm v_{N_r}\}$, respectively. Therefore, the precoders and combiners are defined as
\begin{equation}
\label{deqn_ex3}
\bm u_p=\bm a_t(\phi_p,\frac{\pi}{2}),\qquad p\in\{1,...,N_t\},
\end{equation}
\begin{equation}
\label{deqn_ex4}
\bm v_q=\bm a_r(\phi_q,\frac{\pi}{2}),\qquad q\in\{1,...,N_r\},
\end{equation}
where $\phi_p=\arccos((2p-1-N_t)/{N_t})$, $\phi_q=\arccos((2q-1-N_r)/{N_r})$ $\in[0,\pi)$.

When using the precoder $\bm u_p$ and combiner $\bm v_q$ for beamforming, the received signal strength (RSS), signal-to-noise ratio (SNR), and effective spectral efficiency (ESE) can be described as
\begin{equation}
\label{deqn_ex5}
R_{p,q}=||\sqrt{P_t}\bm v_p^H \bm H \bm u_p s+\bm v_q^H \bm n||^2,
\end{equation}
\begin{equation}
\label{deqn_ex6}
SNR_{p,q}=\frac{||\sqrt{P_t}\bm v_p^H \bm H \bm u_p s||^2}{\sigma _n^2},
\end{equation}
\begin{equation}
\label{deqn_ex12}
ESE_{p,q}=\frac{T_{fr}-N_{b}T_{s}}{T_{fr}}\log_2(1+SNR_{p,q}),
\end{equation}
where $P_t$ denotes the transmitted power, $s$ is the known pilot symbol with normalized power, and $\bm n$ is a zero-mean complex Gaussian noise vector with variance $\sigma _n^2$. Besides, $T_s$ is the time needed to scan a single beam pair, and $T_{fr}$ is the time of a frame with a consistent channel response.  

\section{Location- and Orientation-Assisted Beam Alignment Based on GNN}
In this section, we discuss the correlation between adjacent beams and propose the GNN-based beam alignment method.

\begin{figure}[!t]
\centering
\includegraphics[width=3.4in]{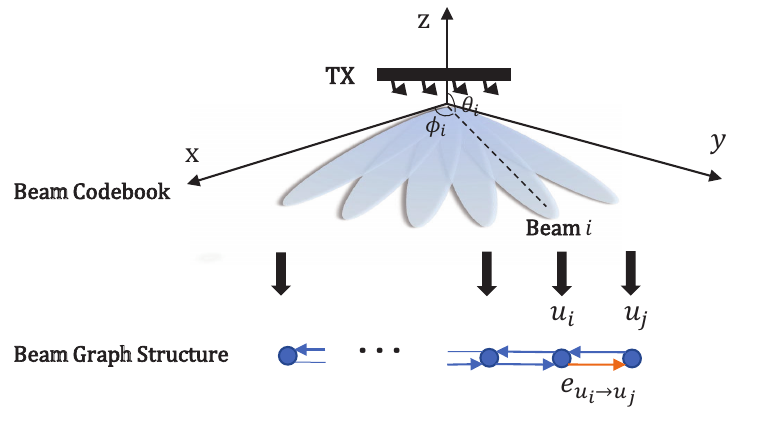}
\vspace{-2ex}
\caption{Beam codebook and beam graph structure for ULA of TX.}
\label{fig_1}
\end{figure}
\subsection{Beam Correlation}
With the known azimuth angles $\phi_p$ and $\phi_q$ and elevation angles $\theta_p$ and $\theta_q$ of two beams, we define the angular correlation $\delta$ between the beams by the cosine similarity, i.e.,
\begin{equation}
\label{deqn_ex7}
\delta=\sin{\theta_p} \sin{\theta_q} \cos{(\phi_p-\phi_q)}+\cos{\theta_p} \cos{\theta_q}.  
\end{equation}

Obviously, the channels of beams are correlated according to the beam direction. When LOS paths exist, the LOS paths of adjacent beams with similar directions are correlated. Moreover, the NLOS paths of adjacent beams are also correlated because the reflection environments of beams with similar directions are likely to be similar. In addition, as the number of antenna elements increases, the beams become narrower, leading to enhanced angular correlation and channel correlation between adjacent beams.

\subsection{Proposed GNN-based Method}
Compared with DNN, GNN has the advantage of good expressive ability due to its capability to capture the graph structural information, which means that GNN can utilize the correlation between beams to effectively select the optimal beam. Besides, since the operations of GNN are defined at the node level and the parameters are shared by all nodes, GNN also has the advantages of few trainable parameters and substantial scalability. 
Therefore, the GNN-based method can be well employed in situations where only a small dataset is available or the beam selection is very complex
, such as in massive multi-input multi-output (MIMO) systems.
\begin{figure*}[!t]
\centering
\includegraphics[width=7.16in]{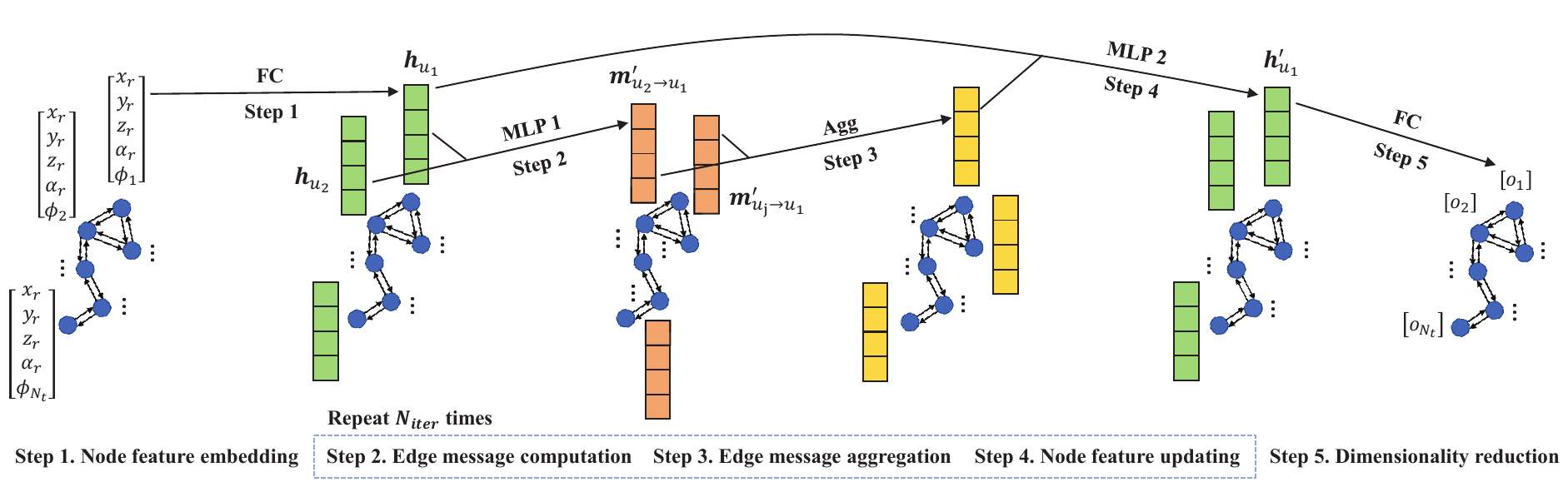}
\vspace{-4ex}
\caption{Proposed GNN architecture using location and orientation for beam alignment.}
\label{fig_2}
\end{figure*}

To leverage the correlation between beams, we define that each node in the graph represents a beam in the codebook, as shown in Fig. 1. Besides, we determine the connection between nodes according to the angular correlation, and the edges are directional. Specifically, a node is connected to the two nodes with the highest correlation with it, and the direction is from the latter to the former. 
For TX, the sets $\mathcal{U}$ and $\mathcal{E}_t$ represent the set of nodes and the set of edges, respectively. Each node $u_i\in{\mathcal{U}}$ has a $n$-dimensional feature vector $\bm h_{u_i}$ and each edge $e_{u_j\rightarrow u_i}\in{\mathcal{E}_t}$ has a $m$-dimensional edge message $\bm m_{u_j\rightarrow u_i}$. The same definition can be extended to RX.

We respectively construct a graph structure for TX and RX and use GNN models for classification. The proposed GNN architecture is depicted in Fig. 2.
In order to deeply utilize the context information, we embed the location $\bm p_r=(x_r,y_r,z_r)$ and orientation $\alpha_r$ of UE as well as the azimuth angle $\phi_i$ of the $i$-th beam in the codebook to obtain the corresponding feature vector $\bm h_{u_i}$ using the fully connected (FC) layer. The initial edge messages are set to zero vectors. The message passing process, i.e., Step 2, Step 3, and Step 4 in Fig.~2, can be represented as 
\begin{equation}
\label{deqn_ex8}
\bm m_{u_j\rightarrow u_i}^{'}=f^{m}(\bm h_{u_j}||\bm h_{u_i}),
\end{equation}
\begin{equation}
\label{deqn_ex9}
\bm h_{u_i}^{'}=f^{n}(\bm h_{u_i}||\underset {u_j \in \mathcal{C}(u_i)} { \operatorname {Agg} } \bm m_{u_j\rightarrow u_i}^{'}),
\end{equation}
where $\bm m_{u_j \rightarrow u_i}^{'}$ is the updated edge message of edge $e_{u_j\rightarrow u_i}$, and $\bm h_{u_i}^{'}$ is the updated feature vector of node $u_i$. $f^{m}$ and $f^{n}$ denote the multi-layer perceptrons (MLPs) with $N_{l,G}$ hidden layers and $N_{h,G}$ neurons per layer. Furthermore, $\mathcal{C}(u_i)$ is the set of nodes pointing to node $u_i$, $||$ denotes the operation of concatenating vectors, and Agg denotes the message aggregation function. 

Obviously, supplementary information about the environment and channel can be obtained from the beam features of adjacent directions through this process, thereby enabling a more effective judgment of whether the current beam is optimal. 

After repeating the message passing process $N_{iter}$ times, the feature vectors of nodes are dimensionally reduced and input into a Softmax function to obtain the optimal probabilities of 
beams. By multiplying the optimal probability of each beam of TX with the optimal probability of each beam of RX, we obtain the optimal probability of each beam pair. The optimal probabilities of all beam pairs are then sorted in descending order, and the first $N_b$ beam pairs are selected to form the set of candidate beam pairs $\mathcal{S}$. Finally, we select the optimal beam pair by measuring the RSS of each beam pair in $\mathcal{S}$ using pilots.
Therefore, we define the misalignment probability as 
\begin{equation}
\label{deqn_ex11}
P(\mathcal{S})=\mathbb{P}[\underset {(i,j)\in \mathcal{S}} { \operatorname {argmax}}R_{i,j}\ne \underset {(p,q)\in \mathcal{B}} { \operatorname {argmax}}R_{p,q}],
\end{equation}
where $\mathcal{B}$ is the set of all possible beam pairs. 

During training, we label the outputs of training data $\mathcal{O}_t=\{o_i|i=1,...,N_t\}$, $\mathcal{O}_r=\{o_j|j=1,...,N_r\}$ for the TX and RX, respectively, as
\begin{equation}
\label{deqn_ex10}
o_i,o_j=
\begin{cases} 
1,  & \text{if $(i,j)=\underset {(p,q)\in \mathcal{B}} { \operatorname {argmax}}R_{p,q}$,} \\
0, & \text{otherwise.}
\end{cases}
\end{equation}

The labels are then converted into one-hot vectors in the computation of the cross-entropy as loss function. 

\section{Simulation Results and Analysis}
In this section, we numerically compare the GNN-based beam alignment with the DNN-based method in \cite{ref4} and the DNN-based method with TL in \cite{ref6}. Besides, we use the same simulation scenario and dataset as \cite{ref6}  for a fair comparison.
In addition to the DNN-based methods as baselines, we use perfect alignment as an optimal solution, which always maintains 100\% alignment accuracy under different numbers of beam pairs scanned. In the simulation, the perfect alignment provides the upper bound for ESE, and the exhaustive search provides the upper bound for RSS.

\subsection{Simulation Setup}
The simulation scenario we considered is the living room (LR) scenario detailed in the IEEE 802.11ad task group, with dimensions of 7m$\times$7m$\times$3m \cite{ref9}. As shown in Fig. 3, the TX is located at the position of (0, 0, 0), and the RX is positioned within the range of $([1.5, 5.5], [-3.5, 3.5], [0])$. Besides, there are five stationary obstacles in the scenario, including two sofas, one table, one chair, and one cabinet. 

We assume the TX and RX operating at 60 GHz, with a bandwidth of 1 GHz and an SNR threshold for scanning of $SNR_{TH}=10$ dB. The parameters used for simulation are $N_t = 64$, $N_r = 16$, $P_t = 0$ dBm, $\sigma _n^2 = -84$ dBm, $T_{fr} = 20$ ms, and $T_s = 0.1$ ms. Using the ray tracing tool, Altair Feko-Winprop, we obtain the channel information of 20 paths of 70,000 location points in the scenario. According to \eqref{deqn_ex1}, the channel matrix can be constructed. Then, according to \eqref{deqn_ex5} and \eqref{deqn_ex6}, the RSS and SNR of each beam pair at each location can be calculated to obtain the dataset. For evaluation, we randomly divide the dataset into a training set and a test set according to 8:2. 
\begin{figure*}[t]
	\centering
	\begin{minipage} {0.32 \linewidth}
		\centering
		\vspace{4ex}
	    \includegraphics[width=0.9\textwidth]{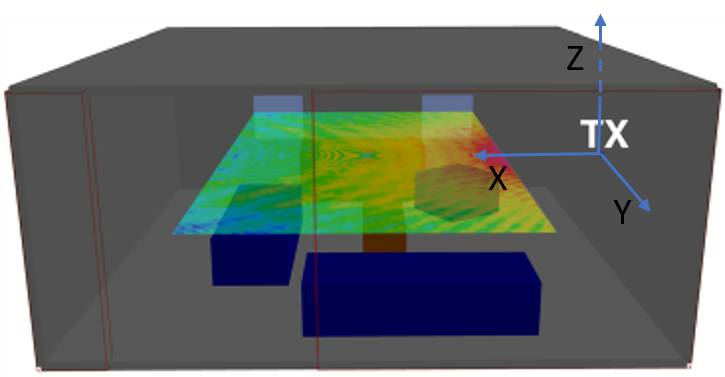}
	    \vspace{4ex}
	    \caption{The ray-tracing simulation of the LR  \protect\\  scenario.}
	    \label{fig_2}
	\end{minipage}
	\begin{minipage} {0.32 \linewidth}
		\centering
	    \includegraphics[width=1\textwidth]{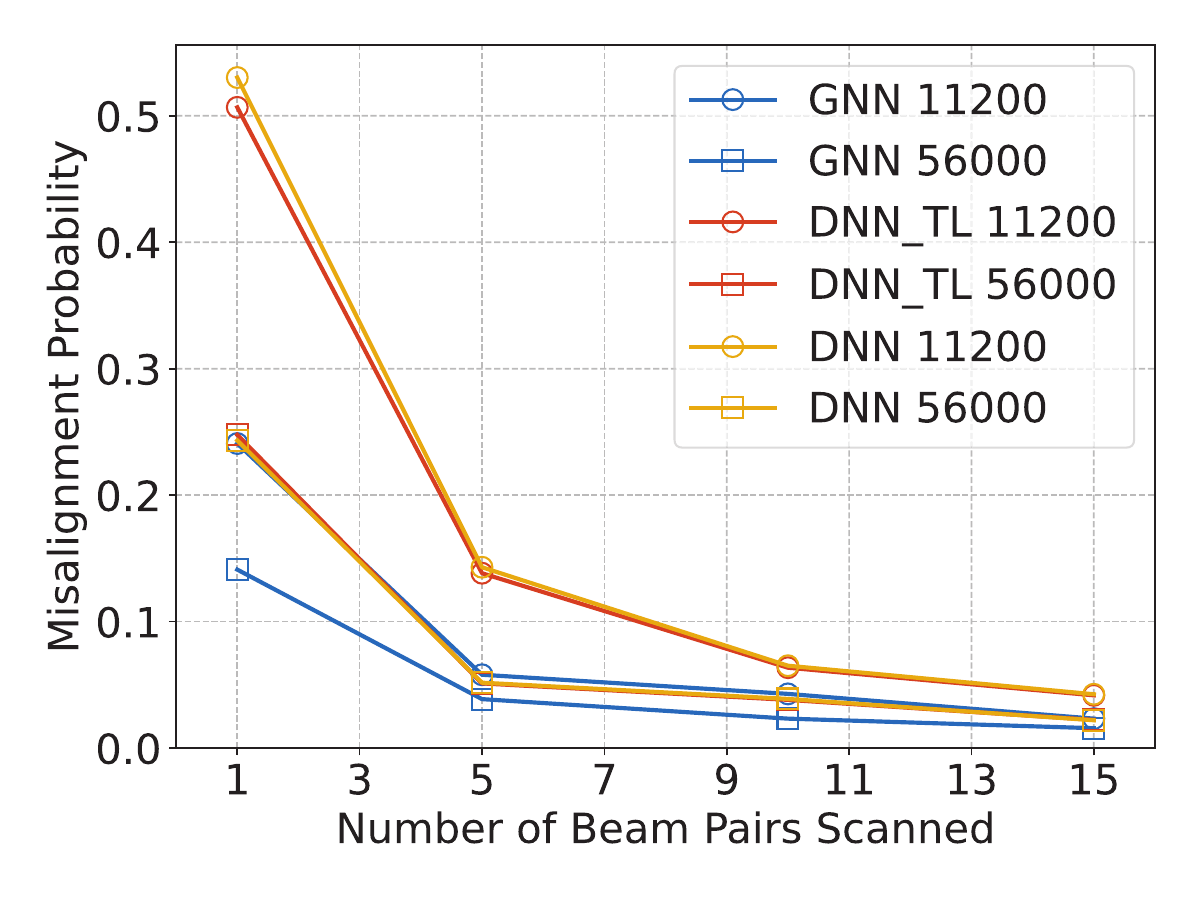}
	    \vspace{-5ex}
	    \caption{Misalignment probability for different \protect\\  training dataset sizes.}
	    \label{fig_3}
	\end{minipage}
	\begin{minipage} {0.32 \linewidth}
		\centering
	    \includegraphics[width=1\textwidth]{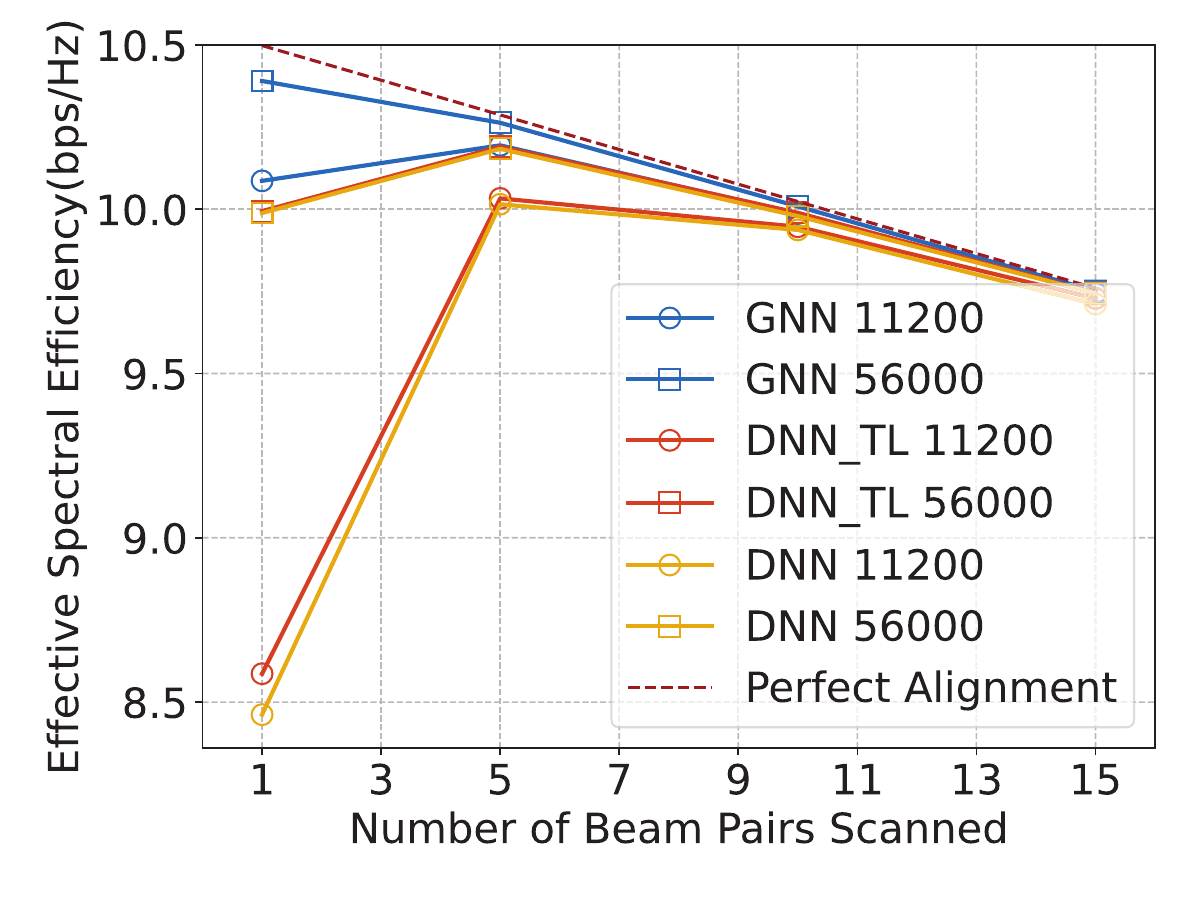}
	    \vspace{-5ex}
	    \caption{Effective spectral efficiency for different \protect\\  training dataset sizes.}
	    \label{fig_2}
	\end{minipage}
	\vspace{-2ex}
\end{figure*}

\begin{figure*}[t]
	\centering
	\begin{minipage} {0.32 \linewidth}
		\centering
	    \includegraphics[width=1\textwidth]{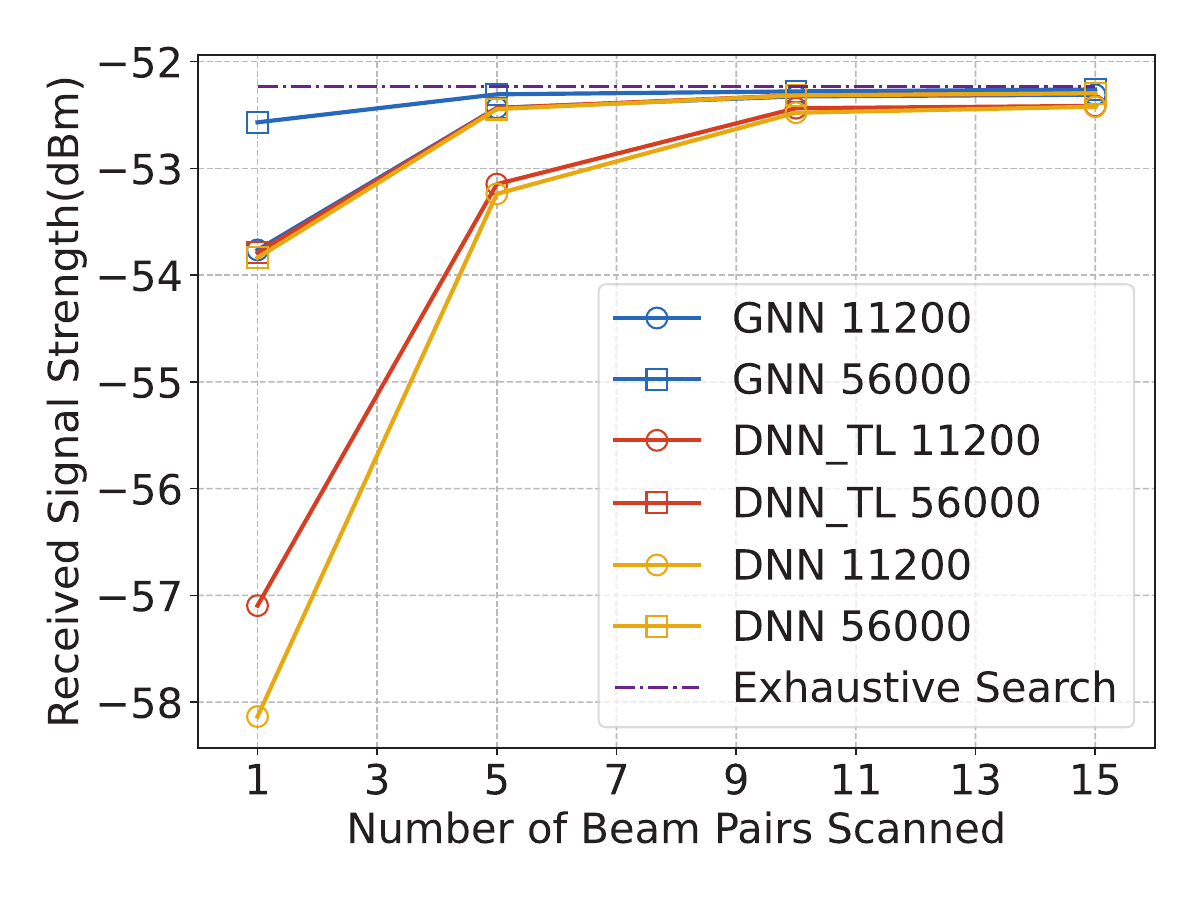}
	    \vspace{-5ex}
	    \caption{Received signal strength for different \protect\\  training dataset sizes.}
	    \label{fig_2}
	\end{minipage}
	\begin{minipage} {0.32 \linewidth}
		\centering
	    \includegraphics[width=1\textwidth]{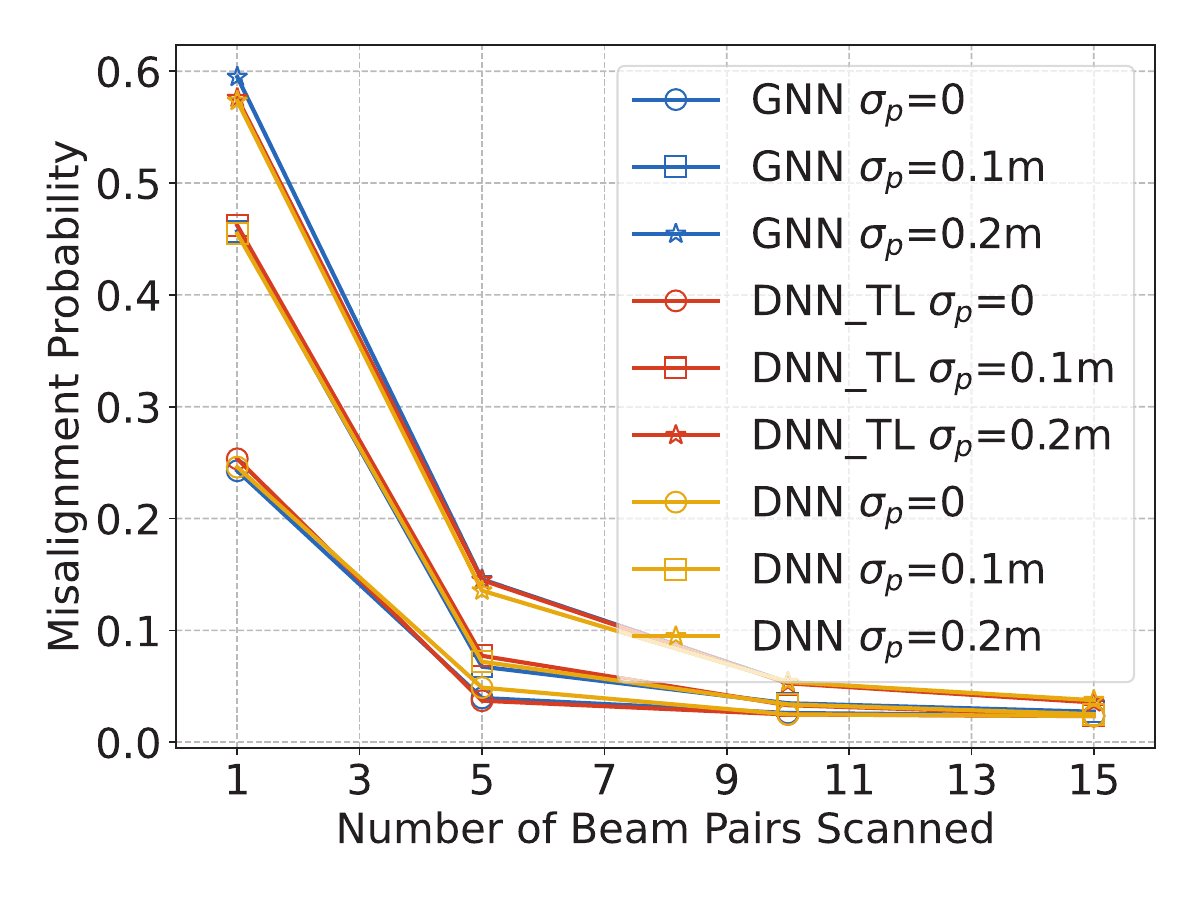}
	    \vspace{-5ex}
	    \caption{Misalignment probability for different \protect\\  location errors.}
	    \label{fig_3}
	\end{minipage}
	\begin{minipage} {0.32 \linewidth}
		\centering
	    \includegraphics[width=1\textwidth]{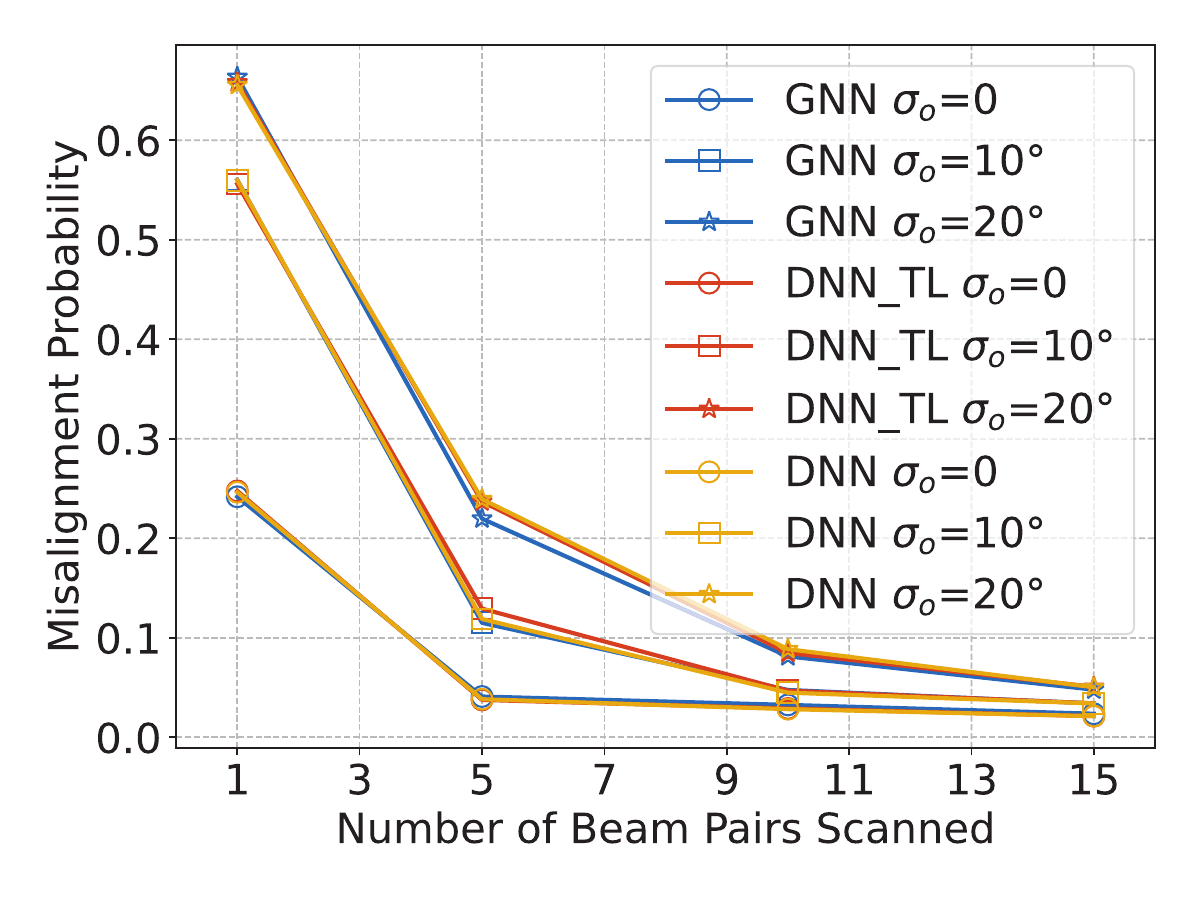}
	    \vspace{-5ex}
	    \caption{Misalignment probability for different \protect\\  orientation errors.}
	    \label{fig_2}
	\end{minipage}
\end{figure*}

Based on empirical observations, we determine the parameters of GNN as: $n=m=16$, $N_{iter}=1$, $N_{l,G}=1$, $N_{h,G}=32$, and choose sum as the message aggregation function. Besides, the DNN-based model is set to have $N_{l,D}=3$ hidden layers with $N_{h,D}=256$ neurons per layer according to \cite{ref4,ref6} and our test results. Also, we adopt the Adam optimizer \cite{ref10} with the learning rate being $10^{-3}$ and the batch size being 128, and the stop criterion of early stopping is adopted.

\subsection{Numerical Evaluation}
Using a subset of the training set, we evaluate the impacts of training dataset size on model performance. Fig. 4, Fig. 5, and Fig. 6 show the misalignment probability, ESE, and RSS of the investigated beam alignment methods with training dataset sizes of 11,200 and 56,000, respectively. Evidently, under different training dataset sizes, the GNN-based method outperforms the DNN-based methods in terms of misalignment probability, ESE and RSS. Besides, the smaller the training set size is, the more obvious the performance improvement would be. When the training set size is 11,200, the proposed method achieves comparable performance to the DNN-based methods with a training set size of 56,000, i.e. the required training set size is reduced to 20\%. This fully demonstrates the advantage of the GNN-based method that aggregates the beam features of adjacent directions to obtain supplementary information about the environment and channel. Additionally, with a training set size of 56,000, the proposed GNN-based method improves the accuracy by 10\% compared with the DNN-based methods when scanning one beam pair, and achieves similar accuracy and RSS to the exhaustive search when scanning 5 beam pairs. Besides, as shown in Fig. 4 and Fig. 6, as the number of scanned beam pairs increases, the misalignment probability decreases, resulting in an increase in RSS. However, the increase in the number of scanned beam pairs also extends the scanning time. Both scanning time and RSS are critical factors influencing the ESE. As shown in Fig. 5, a modest increase in the number of scanned beam pairs can enhance the ESE, but an excessive number of scanning beams can lead to a decrease in that. This suggests an optimal point where the benefits of improved signal strength outweigh the penalties of increased scanning time.

Due to noise and interference, the estimated location and orientation of UE may be inaccurate and the estimation error further influences the accuracy of beam alignment. To evaluate the robustness of models, we add random perturbations to the location $\bm p_r=(x_r,y_r,z_r)$ and orientation $\alpha_r$, as
\begin{equation}
\label{deqn_ex13}
\bm p_r^{'}=\bm p_r+\bm\varepsilon, \qquad \alpha_r^{'}=\alpha_r+\epsilon,
\end{equation}
where $\bm\varepsilon$ is a zero-mean Gaussian random vector with variance of $\sigma _p^2$, and $\epsilon$ is a zero-mean Gaussian random variable with variance of $\sigma _o^2$. 

When the GNN-based method uses a training set size of 11,200 and the DNN-based methods use a training set size of 56,000, the misalignment probability with inaccurate locations and orientations are depicted in Fig. 7 and Fig. 8. Evidently, the proposed GNN-based method not only reduces the required training set size to 20\%, but also exhibits comparable robustness against the DNN-based methods when trained on the reduced dataset. This showcases that leveraging the correlation between beams also enhances performance stability.

Fig. 9 and Fig. 10 illustrate the Top-1 misalignment probability for different numbers of antennas at the TX and RX. It can be observed that the Top-1 misalignment probability of the DNN-based methods increases rapidly with the increase of antenna elements, while that of the proposed GNN-based method is more stable. This indicates that, as the number of antennas increases, improvement by utilizing the channel correlation between adjacent beams is more remarkable. 
Therefore, in massive MIMO systems, the proposed GNN-based method has significant advantages.

\subsection{Complexity Analysis}
For the inference process, the number of multiplications of the GNN-based method and the DNN-based method is $(N_t+N_r)(6F_n+N_{iter}(6F_nN_{h,G}+3(N_{l,G}-1)N_{h,G}^2+3F_mN_{h,G}))=376,320$ and $4N_{h,D}+(N_{l,D}-1)N_{h,D}^2+N_{h,D}N_tN_r=394,240$, respectively. This result verifies that the computational complexity of the proposed method is 4.5\% less than that of the DNN-based method. Meanwhile, the number of trainable parameters of the proposed GNN-based method is $2(6F_n+4F_nN_{h,G}+2(N_{l,G}-1)N_{h,G}^2+2F_mN_{h,G})=$6,336, while that of the DNN-based method is 394,240. Since the computational complexity of one sample in the training process is positively related to the number of trainable parameters, the GNN-based method has a much lower training complexity.

\begin{figure*}[t]
	\centering
    \begin{minipage} {0.32 \linewidth}
		\centering
	    \includegraphics[width=1\textwidth]{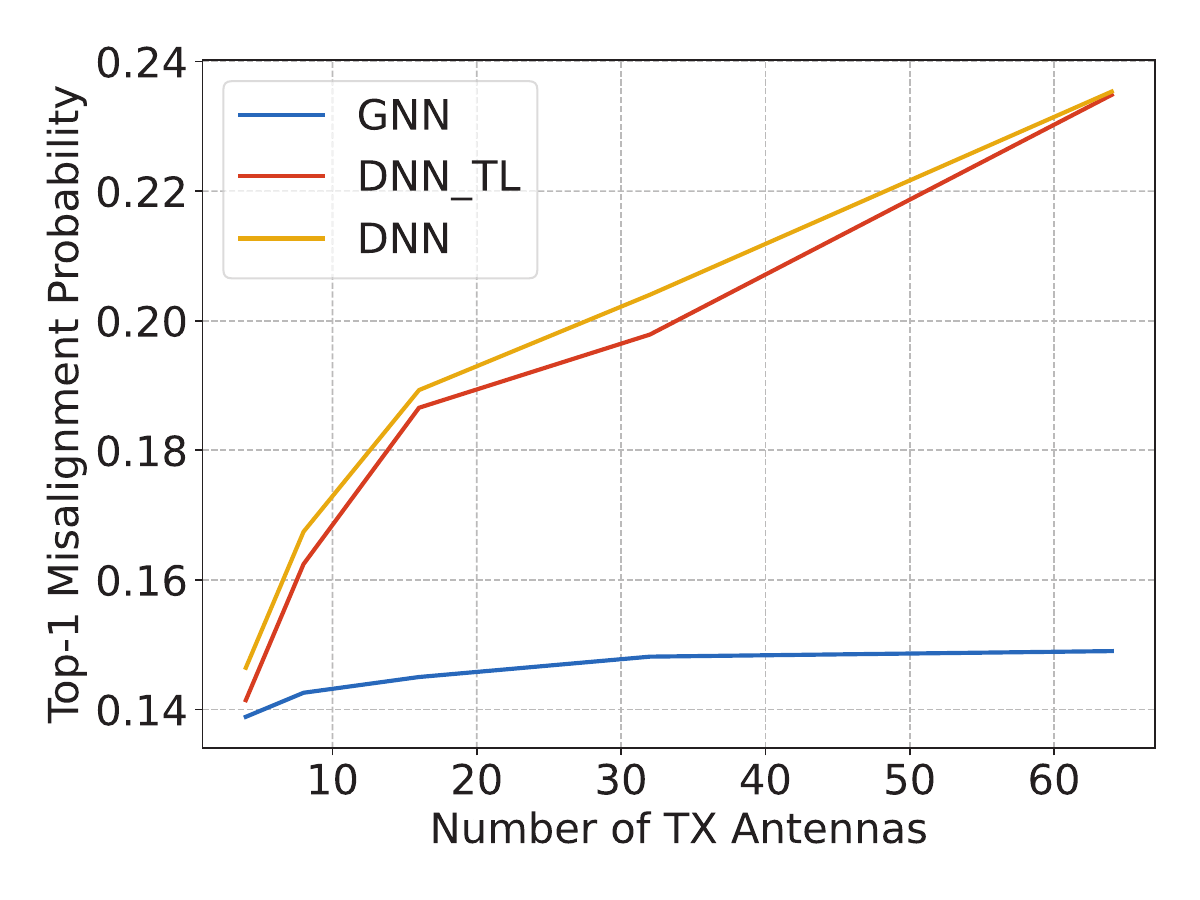}
	    \vspace{-5ex}
	    \caption{Top-1 misalignment probability for \protect\\  different $N_t$ with $N_r=16$.}
	    \label{fig_3}
	\end{minipage}
	\begin{minipage} {0.32 \linewidth}
		\centering
	    \includegraphics[width=1\textwidth]{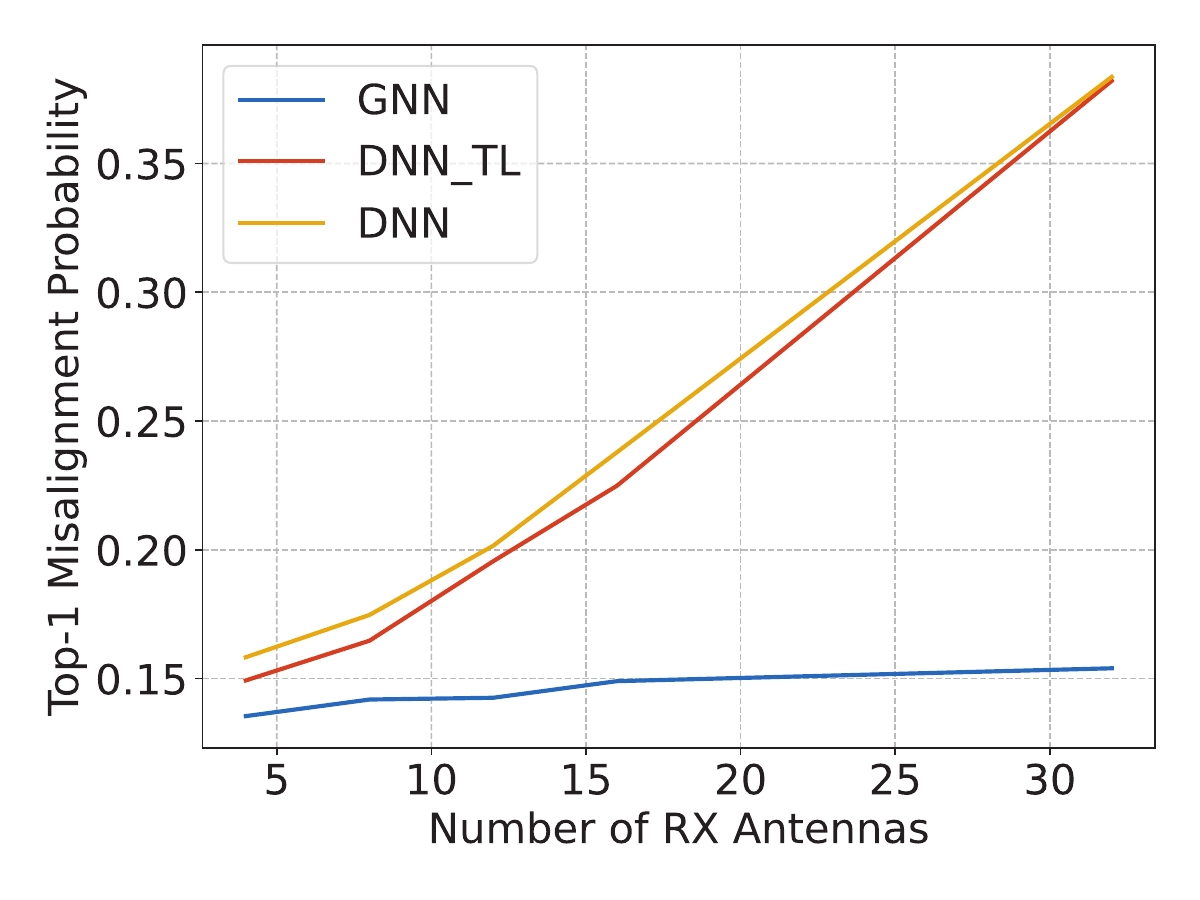}
	    \vspace{-5ex}
	    \caption{Top-1 misalignment probability for \protect\\  different $N_r$ with $N_t=64$.}
	    \label{fig_2}
	\end{minipage}
	\begin{minipage} {0.32 \linewidth}
		\centering
	    \includegraphics[width=1\textwidth]{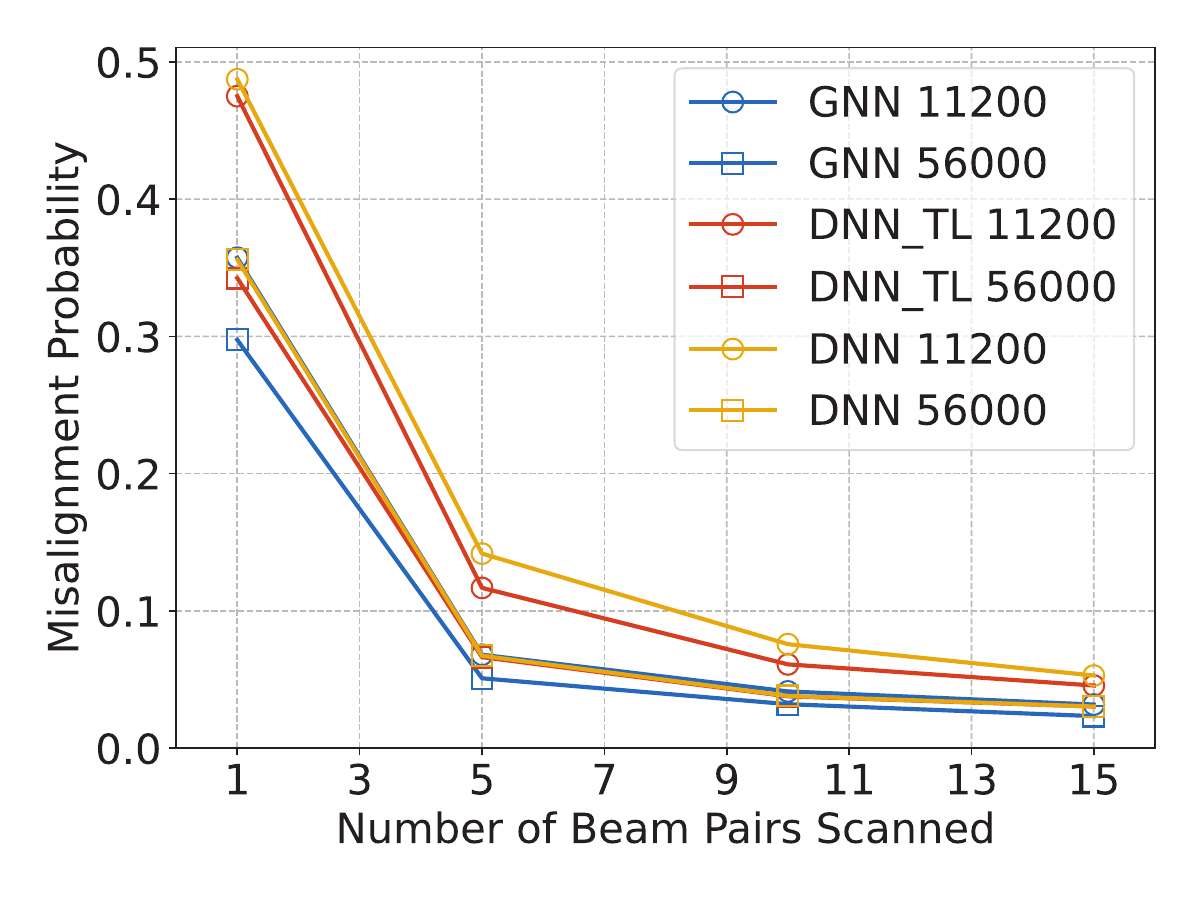}
	    \vspace{-5ex}
	    \caption{Misalignment probability for different \protect\\  training dataset sizes for the UPA scenario.}
	    \label{fig_2}
	\end{minipage}
\end{figure*}

\subsection{Extension to the UPA Scenario}
We extend the proposed method to the scenario where TX and RX are equipped with uniform planar arrays (UPAs) with $\{N_{th},N_{tv}\}$ and $\{N_{rh},N_{rv}\}$ elements separated by $\lambda/2$ in horizontal and vertical dimensions, respectively. To simulate the rotation of UE, we consider the antenna array of RX rotates around z, y, x axes, with the orientation provided by the rotation angles. 

The TX is located at $\bm p_t=(0,0,0)$ with a fixed orientation $\bm \psi_t=(\alpha_t,\beta_t,\gamma_t)$, and the RX is located at a random location $\bm p_r=(x_r,y_r,z_r)$ within the service area $([1.5, 5.5], [-3.5, 3.5], [-0.5, 1])$ with a random orientation $\bm \psi_r=(\alpha_r,\beta_r,\gamma_r)$ and $\alpha_r\in[0,2\pi), \beta_r\in[-\pi⁄4,\pi⁄4), \gamma_r\in[-\pi⁄4,\pi⁄4)$. 

The antenna array response of a UPA with $\{N_h,N_v\}$ antennas can be described as
\begin{align}
\bm a(N_h,&N_v,\phi,\theta)=\frac{1}{\sqrt{N_hN_v}}[1,...,e^{j\pi[ i\sin(\theta)\cos(\phi)+j\cos(\theta)]},\nonumber\\
&...,e^{j\pi[(N_h-1) \sin(\theta)\cos(\phi)+(N_v-1)\cos(\theta)]}]^T, \label{eqn-14}
\end{align}
where $i\in\{1,...,N_h\}$ and $j\in\{1,...,N_v\}$.
Thus we have $\bm a_t(\phi_{t,l},\theta_{t,l})=\bm a(N_{th},N_{tv},\phi_{t,l},\theta_{t,l})$ and $\bm a_r(\phi_{r,l},\theta_{r,l})=\bm a(N_{rh},N_{rv},\phi_{r,l},\theta_{r,l})$ in the UPA scenario.

Using the DFT codebooks, the precoders and combiners in the UPA scenario are
\begin{equation}
\label{deqn_ex3}
\bm u_{a,b}=\bm a_t(\phi_a,\theta_b),  \ a\in\{1,...,N_{th}\},  \ b\in\{1,...,N_{tv}\},
\end{equation}
\vspace{-2ex}
\begin{equation}
\label{deqn_ex4}
\bm v_{c,d}=\bm a_r(\phi_c,\theta_d), \ c\in\{1,...,N_{rh}\},  \ d\in\{1,...,N_{rv}\},
\end{equation}
where $\phi_a=\arccos((2a-1-N_{th})/{N_{th}})$, $\theta_b=\arccos((2b-1-N_{tv})/{N_{tv}})$, $\phi_c=\arccos((2c-1-N_{rh})/{N_{rh}})$, $\theta_d=\arccos((2d-1-N_{rv})/{N_{rv}})$ $\in[0,\pi)$. Mapping all tuples $(a, b)$ and $(c, d)$ to the sets $\{1, 2,...,N_{th}N_{tv}\}$ and $\{1, 2,...,N_{rh}N_{rv}\}$, we obtain the codebooks of TX and RX.

We establish the graph structure according to the angular correlation between beams and take the location $\bm p_r$ and orientation $\bm \psi_t$ of UE as well as the azimuth angles and elevation angles of beams as inputs. Then, we utilize the same GNN architecture for message passing and information utilization to select the optimal beam pair. The parameters used for simulation are $N_{th}=N_{tv}=8$, $N_{rh}=N_{rv}=4$. 

Fig. 11 demonstrates the misalignment probability of investigated beam alignment methods with training dataset sizes of 11,200 and 56,000. In comparison to the DNN-based methods, the proposed GNN-based method also reduces the required training samples to 20\% and achieves higher accuracy when using the same dataset in the UPA scenario.

\section{Conclusion}
The finding of this paper shows that the proposed GNN-based location- and orientation-assisted beam alignment reduces the required training dataset size to 20\%, and improves the Top-1 accuracy by 10\% with the same dataset when compared to the DNN-based methods. Additionally, the proposed scheme exhibits robustness to location and orientation errors and stability for antenna arrays of different sizes. Future investigations will center on more complex communication scenarios, such as designing the extension of the proposed method in orthogonal frequency division multiplexing (OFDM) system.

\end{document}